\documentclass[onecolumn]{aastex631}

\usepackage{amsmath}
\usepackage{comment}
\usepackage{wasysym}

\begin{document}

\title{Effect of multiple scattering on the Transmission spectra and the Polarization phase curves for Earth-like Exoplanets}

\author[0000-0002-2858-9385]{Manika Singla}
\affiliation{Indian Institute of Astrophysics, Koramangala 2nd block, Bengaluru 560034}
\affiliation{Department of Physics, Pondicherry University, Puducherry 605014}

\author[0000-0001-6703-0798]{Aritra Chakrabarty}
\affiliation{Data Observatory Foundation, DO;  Diagonal Las Torres N$^\circ$2640, Building E, Pe\~{n}alol\'{e}n, Santiago, Chile}
\affiliation{Facultad de Ingeniería y Ciencias, Universidad Adolfo Ib\'a\~nez, Av.\ Diagonal las Torres 2640, Pe\~nalol\'en, Santiago, Chile}
\affiliation{Millennium Institute for Astrophysics, Chile}

\author[0000-0002-6176-3816]{Sujan Sengupta}
\affiliation{Indian Institute of Astrophysics, Koramangala 2nd block, Bengaluru 560034}


\begin{abstract}
It is the most appropriate time to characterize the Earth-like exoplanets in order to detect biosignature beyond the Earth because such  exoplanets will be the prime targets of big-budget missions like JWST, Roman Space Telescope, HabEx, LUVOIR, TMT, ELT, etc. We provide models for the transmission spectra of the Earth-like exoplanets by incorporating effects of multiple scattering. For this purpose we numerically solve the full multiple-scattering radiative transfer equations instead of using Beer-Bouguer-Lambert’s law  that doesn't include the  diffuse radiation due to scattering. Our models demonstrate that the effect of this diffuse transmission radiation can be observationally significant, especially in the presence of clouds. We also calculate the reflection spectra and polarization phase curves of  {Earth-like} exoplanets by considering both cloud-free and cloudy atmospheres. We solve the 3D vector radiative transfer equations numerically and calculate the phase curves of albedo and disk-integrated polarization by using appropriate scattering phase matrices and integrating the local Stokes vectors over the illuminated part of the disks along the line of sight. We present the effects of the globally averaged surface albedo on the reflection spectra and phase curves as the surface features of such planets are known to significantly dictate the nature of these observational quantities. Synergic observations of the spectra and phase curves will certainly prove to be useful in extracting more information and reducing the degeneracy among the estimated parameters of terrestrial exoplanets. Thus, our models will play a pivotal role in driving future observations.
\end{abstract}

\keywords{planets and satellites: atmospheres --- atmospheric effects --- transmission spectroscopy --- reflection spectroscopy --- radiative transfer --- polarization --- scattering}

\section{Introduction} \label{sec:intro}

Over 5000 extra-solar planets have been detected till date and many techniques are being developed to study their atmospheres in detail. Techniques such as reflection, transmission and emission photometry and spectroscopy\citep{tinetti2006characterizing, seager2010exoplanet} help in characterizing the planetary atmospheres. Characterizing the terrestrial exoplanets is extremely challenging because of their very small size and low planet to star flux ratio \citep{selsis2008terrestrial, rice2014detection}. We will only be able to detect bio-signature on extra-terrestrial planets unambiguously if we can precisely characterize the Earth-sized planets which are in the circumstellar habitable zone of their host stars \citep{huang1959occurrence, huang1960life, whitmire1991habitable, kasting1993habitable, kopparapu2013habitable, torres2015validation, kane2016catalog, fujii2018exoplanet, covone2021efficiency}. Presence of the  {biosignatures} like oxygen, water, methane, etc. signals the high chances of the presence of life on these planet. The presence of oxygen and ozone is the result of an extended biomass production through oxygenic photosynthesis \citep{owen1980search, sagan1993search,selsis2002signature, selsis2004atmosphericbiomarkers, segura2007abiotic, seager2008exoplanet, selsis2008terrestrial, scharf2009extrasolar, grenfell2014sensitivity, fujii2018exoplanet, claudi2019biosignatures}.

When an exoplanet transits the host star, a fraction of the starlight passes through the planetary atmosphere. The radiation interacts with the atmosphere through scattering and absorption and provides the spectral fingerprints on the transmitted flux. Model transmission spectra for terrestrial exoplanets have previously been presented by \cite{ehrenreich2006transmission, kaltenegger2009transits, palle2009earth, palle2010observations, yan2015high, wunderlich2019detectability, wunderlich2020distinguishing, lin2021differentiating, gialluca2021characterizing, madden2020high}, etc.  In these models, only the total extinction of the incident stellar flux is considered  by using the Beer-Bouguer-Lambert's law \citep{Tinetti}. These models, albeit include the scattering opacity to the true absorption opacity, do not incorporates the angular distribution of the transmitted photons due to scattering in the planetary atmosphere.  \cite{sengupta2020optical} have considered the in and out scattering for the hot Jupiter like exoplanets while modeling the transmission spectra. In the present work, we have calculated the transmission spectra of the Earth-like planets by solving the multiple-scattering radiative transfer equations by using discrete space theory \citep{peraiah1973numerical}, following \cite{sengupta2020optical}. We demonstrate that diffuse transmission radiation due to scattering can affect the overall broad natures of the transmission spectra, especially when the single-scattering albedo increases in the presence of clouds. 

When incoming stellar flux hits the solid planetary surface, some fraction of it gets reflected, absorbed or transmitted depending on the wavelength and the angle of incidence of the incident stellar radiation \citep{seager2010exoplanet, selsis2008terrestrial}. Study of the reflection spectra and phase curves can add to the information that is obtained from the transmission spectroscopy. Moreover, these techniques can be used to characterize the planets with arbitrary orbital alignment with respect to the line of sight. Previously, \cite{sagan1993search} have obtained the reflection spectra of the Earth using the observations by Galileo satellite. Also, \cite{kawashima2019theoretical, batalha2019exoplanet, segura2005biosignatures, kaltenegger2007spectral, kitzmann2010clouds, kitzmann2010influence, rugheimer2013spectral, rugheimer2018spectra}, among many have calculated the reflected spectra for Earth-like exoplanets and also studied the effects of clouds on the spectra. In this paper, we present new model reflection spectra for the Earth-like exoplanets  for the sake of completeness of our investigation on the atmospheres of these planets. These model spectra have been calculated by using the same numerical method mentioned above.

Several polarimetric techniques are also increasingly being used for the study of exoplanetary atmospheres. Polarimetric studies for planets were initiated with the observation of the Solar system objects and are still being continued \citep[][etc.]{coffeen1969, coffeen1969wavelength, hall1974photometric, michalsky1977whole, west1983photometry, joos2007limb}. \cite{mallama2009characterization} characterized the terrestrial exoplanets based on the phase curves of some solar system planets. \cite{stam2003polarization, stam2003polarizationb, rossi2018pymiedap}, etc. studied the polarization spectra for the extra-solar planets. By studying the polarization profiles, we can extract information about the atmospheric as well as physical properties such as cloud distribution,  mean size of cloud particulate as well as rotation-induced oblateness etc. as demonstrated by \cite{sengupta2001probing, sengupta2006polarization, sengupta2008cloudy, sengupta2009multiple, sengupta2010observed, sengupta2011multiple, sengupta2016detecting, sengupta2016polarimetric} for the case of brown dwarfs and self-luminous exoplanets. In addition, phase dependent polarization of reflected planetary radiation can help understanding more  atmospheric composition including  {biosignatures}, surface constitutents like ocean, ice, forest etc and thus the evidence of a habitable environment in exoplanets \citep{zubko2008polarization, kedziora2010using, rossi2017retrieving}. The traces of exo-moons are also being searched by means of polarization \citep{sengupta2016detecting, berzosa2017traces, molina2018traces}.

The reflected light can be polarized because of the various scattering processes which depend on the types of scatterers and the scattering mechanism \citep{seager2010exoplanet}. Linear polarization signals from the starlight reflected from the horizontally inhomogenous Earth-like planets is presented in \cite{karalidi2012modeled}. \cite{groot2020colors, rossi2018circular} studied the linearly or circularly polarized signals from the sunlight reflected from the model Earth. Polarization signals from starlight reflected by the Earth-like exoplanets have been studied by \cite{stam2008spectropolarimetric, fauchez2017o2, wei2017polarimetric, munoz2018mapping, sterzik2019spectral, patty2021biosignatures, gordon2022polarized}, among others. \cite{wang2019diurnal} have used PARASOL data to calculate the variation of the disk-integrated polarization. \cite{ karalidi2011flux,karalidi2012looking, michael2020} have modeled the polarized signal from the clouds on exoplanets and \cite{zugger2010light, zugger2011searching} from the exoplanetary oceans and atmospheres. \cite{stam2005errors} have estimated the errors in the calculated phase functions and albedos of planets if polarization is neglected.

Detecting the polarization signals of the reflected radiation is, however, extremely difficult because of the very low signal to noise (S/N) ratio as compared to that of the Solar-system planets.
Some of the upcoming telescopes will unravel the polarimetric properties of the Earth and the extra-solar planets. LOUPE (Lunar Observatory for Unresolved Polarimetry of the Earth), a small spectropolarimeter is being developed to observe the Earth from the Moon as an exoplanet \citep{klindvzic2021loupe, karalidi2012observing} and also ELF (Exo-Life finder) Telescope \citep{berdyugina2018exo} will be used for the direct detection of exoplanet biosignatures. Other big-budget missions like HabEx, LUVOIR, Roman Space Telescope, etc. will also have imaging polarimetric facility. These missions will additionally have coronographic instruments onboard which will allow us to detect polarimetric signals from the exoplanets in the habitable zones directly.

Most of the above mentioned polarization models either use Monte Carlo method or solve 1-D vector radiative transfer equations and invoke generalized spherical harmonic expansion to integrate the scattering polarization over the visible disk.  In the present work we calculate the azimuth-dependent intensity vectors by solving the 3-D vector radiative transfer equations. The disk integrated flux and polarization are estimated by integrating the intensity vector at each local point over the illuminated disk. The 1-D version of the same numerical code has also been used to solve the radiative transfer equations in their vector form for the calculation of polarized spectra of rotation-induced oblate self-luminous exoplanets and cloudy brown dwarfs \citep{sengupta2009multiple, sengupta2010observed, marley2011probing, sengupta2016polarimetric, sengupta2016detecting, sengupta2018polarization}. However, in order to calculate the polarization over the rotation-induced oblate disk of the object, spherical harmonic expansion method was used in those work. This scalar version of the same code has also been used to calculate the transmission spectra for the hot Jupiters \citep{sengupta2020optical, chakrabarty2020effects}. \cite{chakrabarty2021generic} have presented the polarization models for hot-Jupiters by solving 3D vector radiative transfer equations. In the present work we employ the same methodology to calculate the polarization for the Earth-like exoplanets.

In the next section, we discuss about the necessary inputs used to calculate the transmission spectra and the reflected spectra for Earth-like exoplanets.  In sections \ref{sec:trans} and \ref{sec:ref}, we  {present the results of} the transmission and the reflection spectra. Vector phase curve models are presented in  section \ref{sec:vec}. In section \ref{sec:res},  {we analyze and discuss the results} and finally, in the last section, we present the conclusions of this work.

\section{Atmospheric models for Earth-like exoplanets}\label{sec:atmos}

We present models for transmission spectra, reflection spectra and the phase curves for geometric albedo and linear polarization for the Earth-like exoplanets orbiting around Sun-like stars. For calculating the reflection and the transmission spectra as well as the scattering polarization, we take the atmospheric chemical composition  {for the modern Earth-like exoplanets} from \cite{kawashima2019theoretical} and opacity data {, i.e. absorption and scattering cross-sections for all the molecules that have been taken in the atmospheric composition of the planet,}  from the database for PICASO \citep{natasha_batalha_2020_3759675}.
The observed temperature-pressure profile of the Earth's atmosphere is considered for the calculations. We consider two types of atmosphere in all of our model calculations: cloudy and cloud-free. In the case of cloudy atmospheres, we consider very thin clouds or haze and we have used an approximate Rayleigh model to express the effect of these clouds/haze following \cite{sing2016continuum, kempton2017exo}, etc. 

In case of transmission spectra, we have included thin clouds with scattering cross-sections ($\sigma$) equal to  {100, 200 and 400 times} the scattering cross-section ($\sigma_R$) of the dominant atmospheric constituent i.e. nitrogen in this case. The cloud deck and base have been fixed at 5x10$^2$ Pa and 5x10$^3$ Pa. For the case of reflection spectra, we have considered the cloud position between the pressure levels of  {1x10$^3$ Pa and 5x10$^4$ Pa} with a scattering cross-section equal to 400 times the scattering cross-section of nitrogen gas.  {The cloud position (considering 100 \% coverage) for the case of reflected spectra is kept at deeper layers of the atmosphere while for the case of the transmission spectra, the clouds are considered at the upper layers of the atmosphere. It is because we can probe only the outer atmosphere by transmission spectra. Also, we will probe complementary portions of the atmosphere in terms of the altitude.} A terrestrial exoplanet usually should have water cloud in the upper atmosphere. For large cloud particulates, Mie scattering phase matrix should be appropriate to describe the angular distribution of photon before and after scattering. But for small size of cloud particles, Rayleigh phase matrix serves the purpose reasonably well. In the present work we have used  {Rayleigh} phase matrix for water droplets \citep{sengupta2006polarization}.

\section{ {Results}}
\subsection{ {The Transmission Spectra}}\label{sec:trans}

Using the atmospheric models described in Section~\ref{sec:atmos}, we have presented models of transmission spectra of the Earth-like exoplanets. Studying the absorption lines on the transmission spectra can directly tell us about the volatiles and  {biosignatures} present in their atmospheres. But such interpretation requires accurate models of the broad continua of the spectra, especially in the visible wavelengths and in the presence of clouds. Also, an accurate model can help us understand how the presence of clouds can suppress the absorption lines since detecting these absorption lines of the Earth-like planets is already extremely challenging. Following \cite{sengupta2020optical}, we solve the multiple-scattering radiative transfer equation for diffuse reflection and transmission which is given as,
\begin{equation}\label{multscateqn}
\mu\frac{dI(\tau,\mu,\lambda)}{d\tau}=I(\tau,\mu,\lambda)-
\frac{\omega}{2}\int_{-1}^1{p(\mu,\mu')I(\tau,\mu',\lambda)\text{d}\mu'}
-\frac{\omega}{4}F e^{-\tau/\mu_0}p(\mu,\mu_0).
\end{equation}
Here, I($\tau$,$\mu$,$\lambda$) is the specific intensity of the transmitted radiation along our line of sight. $\omega$ is the single scattering albedo, F is the incident stellar flux along our line of sight (along the direction -$\mu_0$), p($\mu$,$\mu'$) is the scattering phase function and $\tau$ is the optical depth along the line of sight. The detail formalism and numerical technique are described in \citep{sengupta2020optical}.

\begin{figure}
\centering
\includegraphics[scale=0.45]{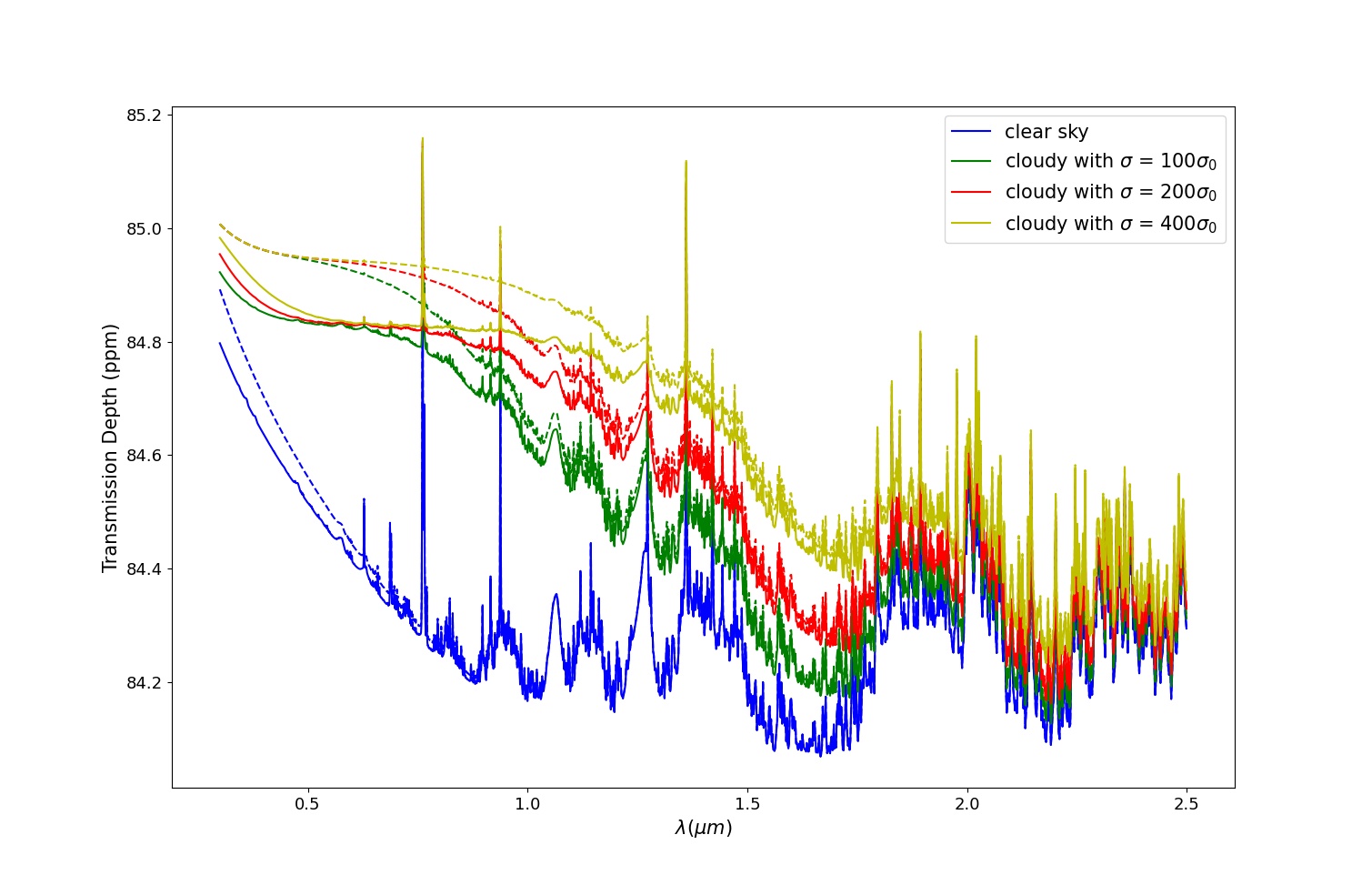}
\caption{Transmission depth of the Earth-like exoplanets with (solid) and without (dashed) diffuse scattering for cloudy and cloud-free atmospheres (see Section~\ref{sec:atmos}). Scattering opacity is however, included even in the case for cloud-free atmosphere.
\label{fig:trans}}
\end{figure}

Surface albedo of the planet is not considered in this case as transmission spectra predominantly convey the information of the upper atmosphere. Figure~\ref{fig:trans} presents the transmission depth with and without diffusion of radiation due to scattering. When the diffusion due to scattering is not considered, especially in the longer wavelengths where the values of $\omega$ are extremely low  {($\omega$ $\approx$ 0)}, we can use the Beer-Bouguer-Lambert's law instead of solving the radiative transfer equations. Note that even when the diffuse radiation due to scattering is not incorporated, total atmospheric optical depth is determined by both the absorption and the scattering opacities. 

\subsection{ {The Reflection Spectra}}\label{sec:ref}

While orbiting the host star, an exoplanet reflects a part of the incident starlight depending on its orbital phase, orbital inclination, and of course, the atmospheric and surface constituents. The study of the reflected light from the planets helps us to probe deeper into their atmospheres, detect their surface as well as cloud properties. It also reduces the degeneracy among the estimated parameters when combined with the results from the study of transmission spectra.

\begin{figure}
\centering
\includegraphics[height=14cm,width=18cm]{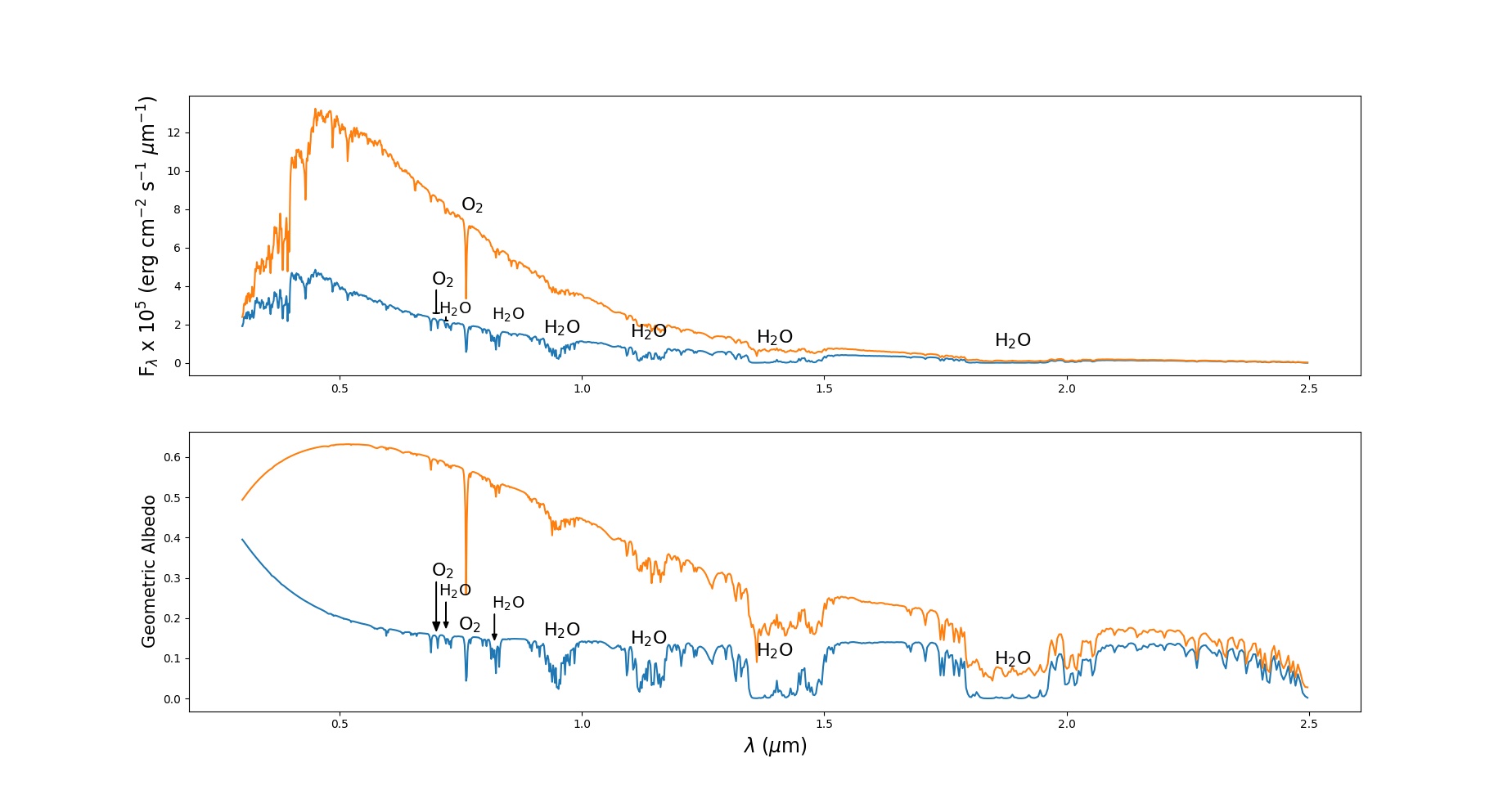}
\caption{Reflected spectra (top panel) and Geometric albedo (bottom panel) for the Earth-like exoplanets orbiting around the Sun-like stars at a resolution of 300. The blue plot is for the clear sky while the orange plot is for the cloudy atmosphere. The absorption lines of O$_2$ and H$_2$O are shown.
\label{fig:reflected}}
\end{figure}

We solve the scalar one-dimensional multiple-scattering radiative transfer equation to model the geometric albedo (see, for example, \cite{batalha2019exoplanet}) and the full-phase (fully illuminated disk) reflection spectra of the Earth-like planets using the atmospheric model explained in Section~\ref{sec:atmos}. The equation is somewhat similar to Equation~\ref{multscateqn} but follows a different geometry (radial geometry) to calculate the layerwise optical depths and single-scattering albedos. Surface albedo, which depends on the surface composition, contributes to the overall reflectivity of the rocky planets. For example, if the whole surface is covered with snow, the surface albedo is the highest, i.e. around 0.9 and hence it contributes heavily to the geometric albedo but if the whole surface is covered with ocean, the surface albedo is much less, around 0.06, thus contributing much less to the geometric albedo \citep{kaltenegger2007spectral}. For the present Earth-like exoplanets, we take the surface albedo to be 0.14 at all wavelengths, where the surface components are 70\% ocean, 2\% coast, and 28\% land, which in turn, is divided into 30\% grass, 30\% trees, 9\% granite, 9\% basalt, 15\% snow, and 7\% sand \citep{kaltenegger2007spectral}. The surface reflection is assumed to be Lambertian i.e. isotropic in nature. We calculate the surface albedo by summing all the components' albedo multiplied by their respective fraction of the planetary surface coverage. Figure~\ref{fig:reflected} shows the reflection spectra and the geometric albedo of cloudy and cloud-free Earth-like planets with the annotations of oxygen and water absorption lines.

\subsection{ {The Phase Curve Models}}\label{sec:vec}

The reflected light observable from the planets depends on their orbital phase and the study of these phase curves conveys valuable information about the atmospheres and surfaces of the Earth-like exoplanets. The orbital phase ($\alpha_{\rm orb}$) is 0$^o$ when the maximum area of the illuminated disk is viewed and 180$^o$ when the minimum or no illuminated part of the planetary disk is viewed. However, modeling these phase curves is cumbersome and requires us to invoke three-dimensional radiative transfer models as explained by \citep{chakrabarty2021generic}. Here, we solve the 3-D vector radiative transfer equation to calculate both the albedo (total reflectivity of the disk) phase curves and the disk-integrated polarization phase curves. 

The partial illumination of a planetary disk yields net non-zero disk-integrated scattering polarization of the reflected light. A study of this polarization can provide us information about the atmospheric clouds in detail, surface composition, and also the light absorbers present in the atmospheres \citep{chakrabarty2021generic}. We assume the incident starlight to be unpolarized and the polarization of the planet's reflected light is solely caused by the scattering process. We ignore polarization due to strong magnetic field if any.

The state of polarization of each beam of ray after scattering is determined by the scattering phase matrices which depend on the scattering mechanism. We follow the methods prescribed by \cite{chakrabarty2021generic} to solve the vector radiative transfer equations and calculate the phase dependent reflected flux and the polarization ($P$) averaged over the illuminated planetary disk. The corresponding atmospheric model is explained in Section~\ref{sec:atmos}. 

\begin{figure}
\centering
\includegraphics[scale=0.5]{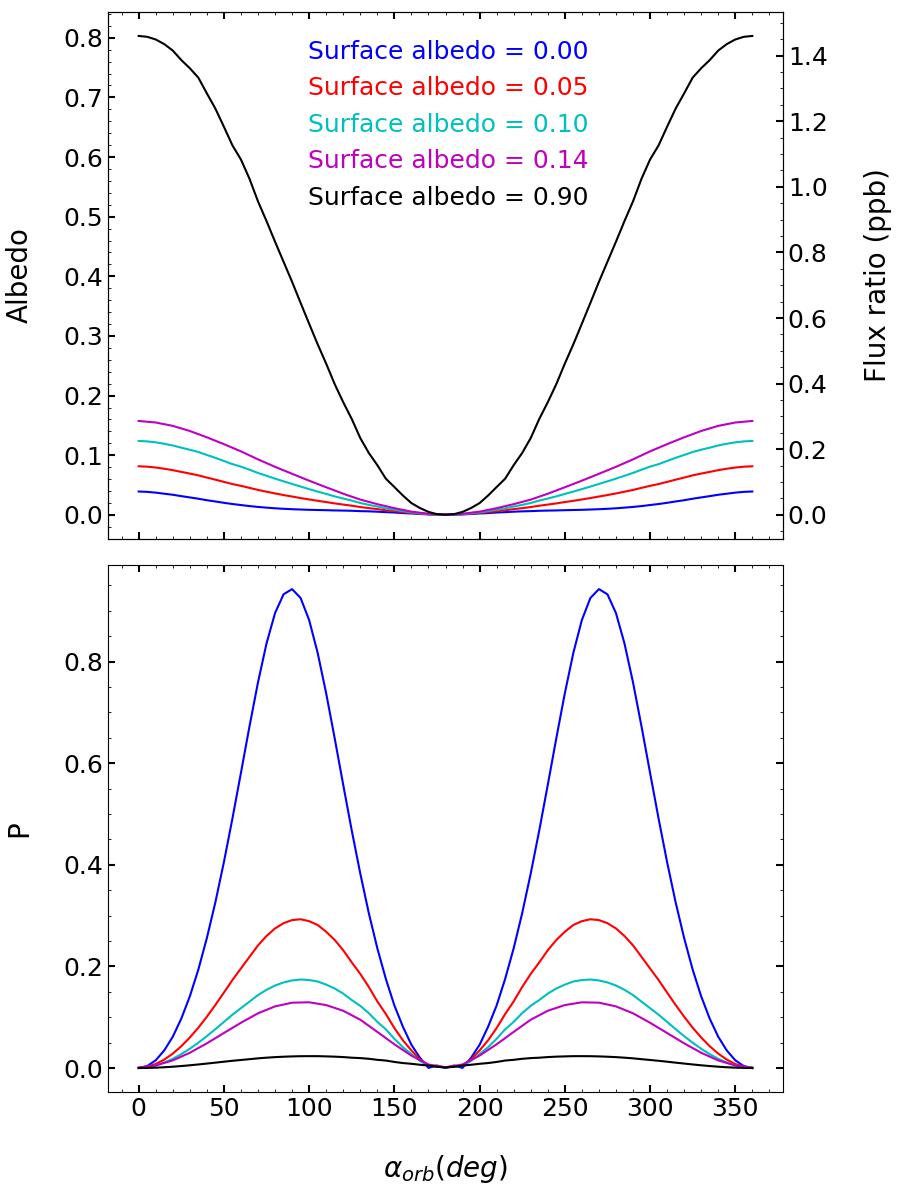}
\caption{Effect of surface albedo on the phase curves of albedo (or flux ratio, F($\alpha_{\rm orb}$)/F$_0$) and net polarization ($P$) integrated over the illuminated planetary disk  at a wavelength of 0.6 $\mu$m and at an orbital inclination = 90$^o$.  {We have used surface albedo 0.14 for our calculations. And 0.9 surface albedo is for the case of snowball (fully covered with snow) planet.}
\label{fig:pc_diff}}
\end{figure}

We studied the effect of surface albedo on the overall disk albedo and polarization as depicted in Figure~\ref{fig:pc_diff}. For the rest of the calculations, we considered the value of surface albedo to be 0.14 as explained in Section~\ref{sec:ref}.

\begin{figure}
\centering
\includegraphics[scale=0.5]{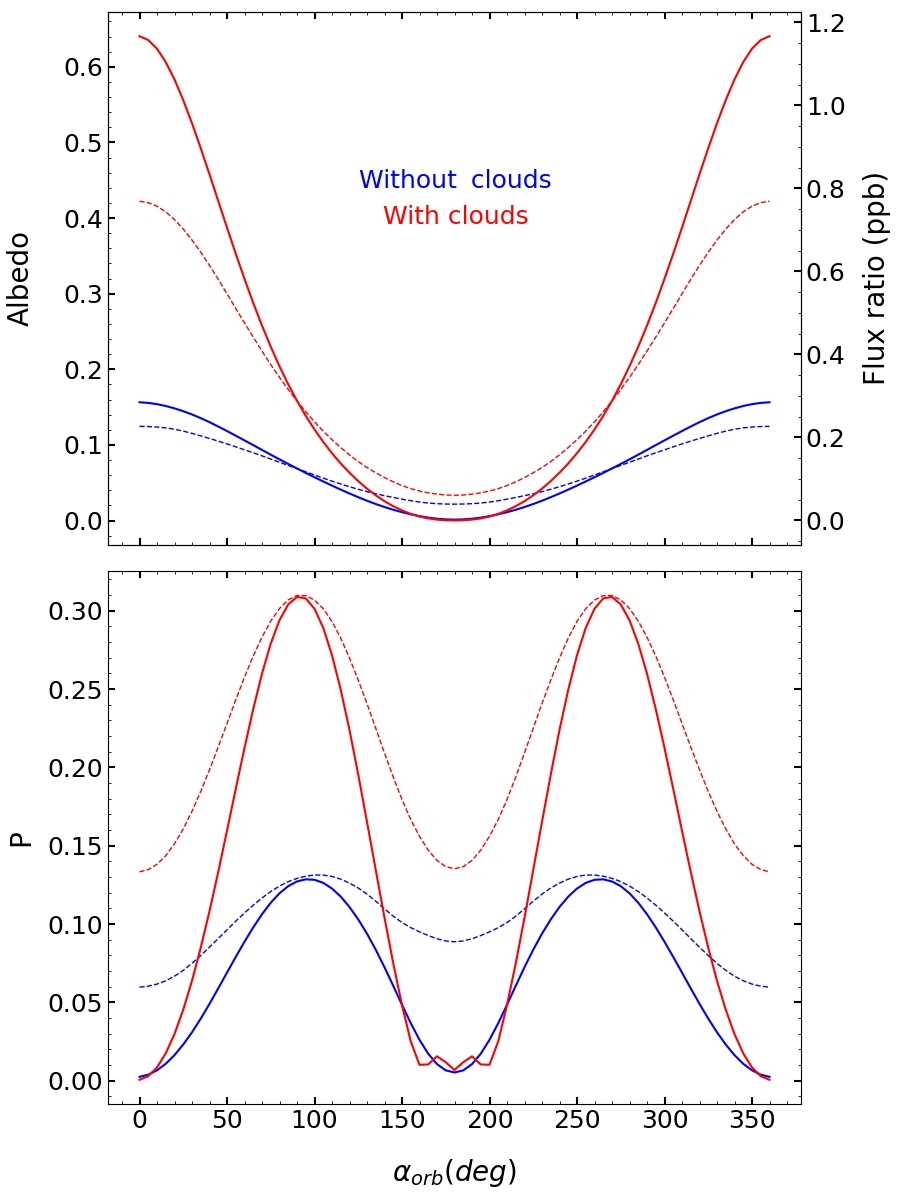}
\caption{The phase curves of albedo (or flux ratio, F($\alpha$)/F$_0$) and  polarization ($P$) integrated over the illuminated disk at a wavelength of 0.6 $\mu$m (visible) and at orbital inclinations angle 90$^o$ (solid) and 45$^o$ (dashed) for both cloud-free and cloudy atmospheres.
\label{fig:pc_vis}}
\end{figure}

\begin{figure}
\centering
\includegraphics[scale=0.5]{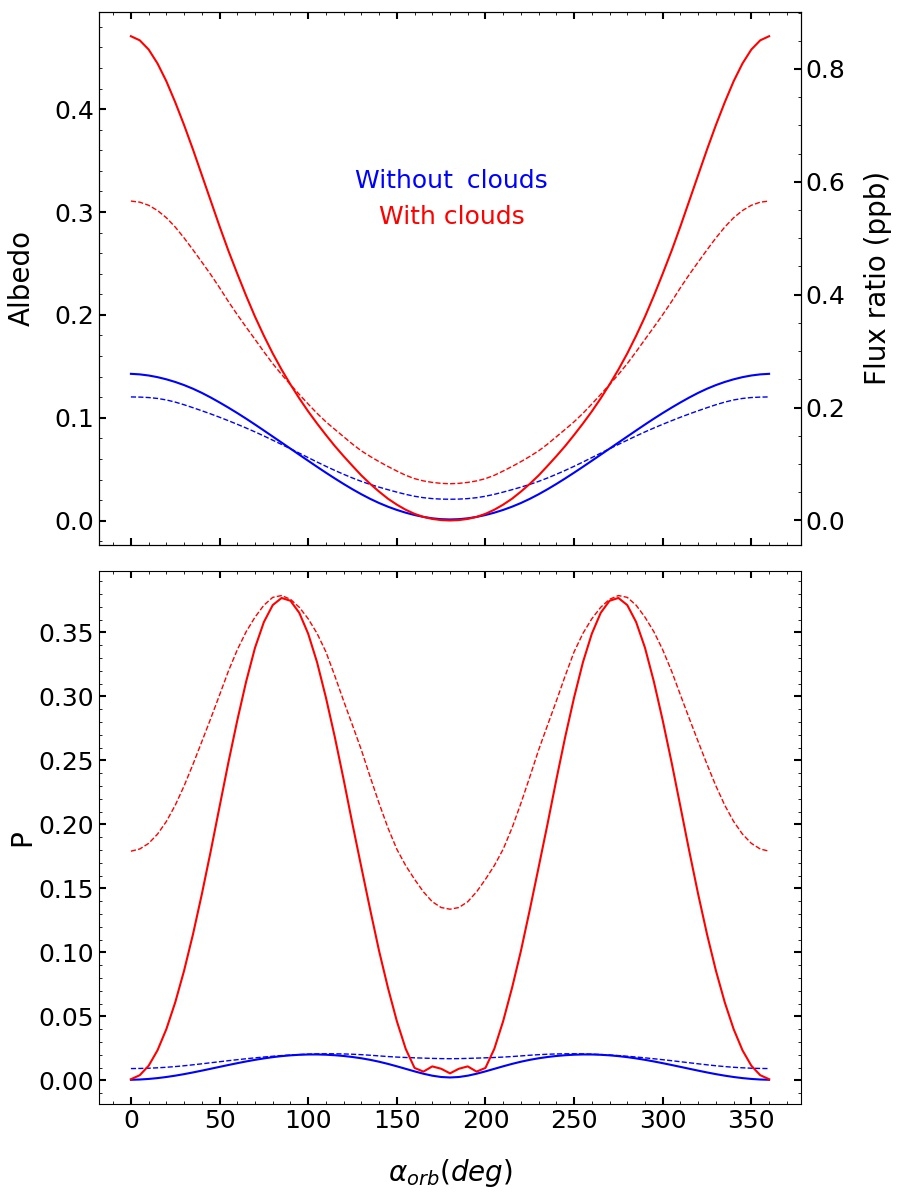}
\caption{Same as figure \ref{fig:pc_vis} but at a wavelength of 1.0 $\mu$m (near infrared).
\label{fig:pc_nir}}
\end{figure}

\begin{figure}
\centering
\includegraphics[scale=0.5]{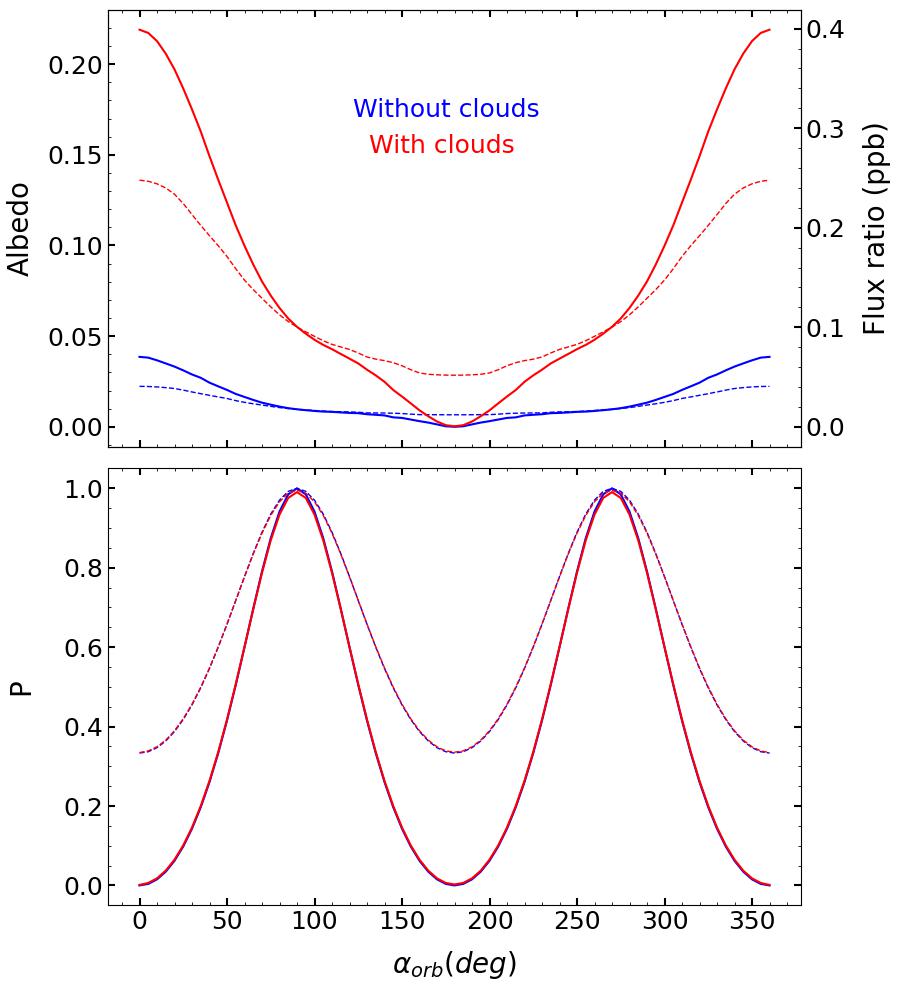}
\caption{Same as figure \ref{fig:pc_vis} but with single scattering of the incident radiation.
\label{fig:pc_single}}
\end{figure}

Figure~\ref{fig:pc_vis}-\ref{fig:pc_nir} show the the total albedo and the disk polarization ($P$) at $\lambda=0.6~\mu$m and $\lambda=1~\mu$m for both the cloud-free and cloudy atmospheres {, by considering multiple scattering of the incident radiation}.
 {Figure \ref{fig:pc_single} shows the same for visible wavelength ($\lambda=0.6~\mu$m), considering only single scattering.}
These phase curves can be detected with the next-generation polarimetric missions which will use their coronagraphic instruments to resolve the Earth-like exoplanets from their host stars. The observable flux ratio i.e. the ratio of the observable reflected flux from the planet to the observable starlight is shown in the figures. This indicates the contrast required by those instruments to directly detect the reflection spectra from such planets. Since the flux ratios are in the order of parts per billion (ppb), detecting the polarization of the planets without resolving them separately amidst the stellar glare will be impossible with the current technology and hence not shown in the figures.

\section{ {Analysis and Discussion}}\label{sec:res}

The transmission depth for modern Earth-like exoplanets orbiting around Sun-like stars at wavelengths up to 2.0 $\mu$m is shown in Figure \ref{fig:trans}. We can see that the transit depth reduces with the inclusion of diffuse radiation due to scattering as explained by \cite{sengupta2020optical}. The transmission depth increases with the inclusion of clouds as the cloud particles block the transmitted flux through the atmosphere. Clouds also suppress the absorption features of the molecules in shorter wavelengths. The effect of diffusion due to scattering on the broadband continuum can be significant with respect to the levels of the individual absorption features, especially in the presence of atmospheric clouds. This calls for more accurate modeling of the transmission spectra by solving the complete radiative transfer equation.  Otherwise detecting the features of the  {biosignatures} of the Earth-like planets can be confusing and may be erroneous. Of course, at longer wavelength regions, all the plots are found to merge because the effect of scattering is negligible.

The relatively stronger absorption lines such as O$_2$, H$_2$O are easily detectable in the reflection spectra (see Figure~\ref{fig:reflected}). Clearly, the geometric albedo increases with the decrease in wavelength because of the dominance of Rayleigh scattering ($\propto \frac{1}{\lambda^4}$) at shorter wavelength. Also, because of increased back scattering of the incident stellar radiation, the presence of clouds significantly increases the geometric albedo and hence the reflected flux at the visible wavelength region. .

Figure \ref{fig:pc_diff} shows the variation of the albedo and the disk-averaged polarization ($P$) for different orbital phase at a fixed wavelength ($\sim$ 0.6$\mu$m) and an orbital inclination angle of 90$^o$ (edge-on view) for different surface albedos. The value of the surface albedo depends on the surface composition of the planet i.e. the amount of ocean cover, land cover, trees, ice, etc.  {The surface albedo of 0.9 corresponds to the case where the whole surface of the planet is covered with snow.} The intermediate surface albedo of 0.05 in the figure corresponds to the case where almost the whole surface is covered with ocean and 0.1 corresponds to the case where half of the surface is covered with ocean and the remaining half with trees and grass. As the surface reflection is assumed to be Lambertian, it completely depolarizes the  light that is reflected in the upward direction from the surface of the planet, i.e., from the bottom of the atmosphere (BOA) \citep{rossi2018pymiedap}. As a result, with an increase in the surface albedo, the overall albedo increases but the polarization ($P$) decreases. $P$ is evidently found to peak at around 90$^o$ orbital phase. The phase dependent polarization profile presented here is consistent with that presented by \citep{sengupta2006polarization, stam2003polarization}.

Figures~\ref{fig:pc_vis} and \ref{fig:pc_nir} show the phase dependent light curves at wavelengths of 0.6 $\mu$m and 1.0 $\mu$m respectively for a planet with two orbital inclinations, e.g., 45$^o$ and 90$^o$ Evidently, the peak-to-peak fluctuations of the light curves decrease with a decrease in the orbital inclination. Moreover, these figures also demonstrate the effects of clouds. The presence of clouds and non-zero surface albedo increase the total albedo of the disk as expected.
 {Figure \ref{fig:pc_single} show the phase dependent light curve at a wavelength of 0.6 $\mu$m for the same orbital inclinations by considering only single scattering at each atmospheric layer.
The effect of clouds on the albedo is the same as the case for multiple scattering. But clouds do not affect the disk averaged polarization because angle of scattering at each layer is the same.
We note that the peak polarization (i.e. at 90$^o$ phase angle) is 1 for clear sky as well as cloudy atmosphere. This happens because we have approximated the effects of clouds with the Rayleigh phase matrix and for the case of Rayleigh scattering, the degree of single-scattering polarization is 1 at 90$^o$ phase angle (see Fig. 3 of \cite{chakrabarty2021generic}). Basically, the single-scattering approximation overestimates the observable polarization and underestimates the total albedo.}

 {Although, we see opposite behaviour of disk averaged polarization with the clouds in case of hot-Jupiters (see Fig. 15 of \cite{chakrabarty2021generic}). For hot-Jupiters, the polarization depends on single scattering albedo, but for the Earth-like exoplanets, it depends on single scattering albedo as well as the surface albedo as explained in the next paragraph.}

However, to understand the total degree of polarization for a planet with a Lambertian surface with non-zero surface albedo, we divide the total upward (towards us) radiation at the top of atmosphere (TOA) into two streams: (i) the downward incident radiations that get scattered back to the upward direction and get polarized, especially at disk locations away from the substellar point \citep[e.g.,][]{stam2006integrating, chakrabarty2021generic}, and (ii) the upward radiations from BOA that get transmitted in the same direction which are predominantly unpolarized. For a cloud-free atmosphere, as the wavelength increases, the intensity of stream-i decreases since the single-scattering albedo of the atmosphere decreases, whereas the intensity of stream-ii remains almost constant as we have assumed the same surface albedo at both the wavelengths. Hence, the relative dominance of stream-ii increases at higher wavelengths, and the polarization ($P$) drops significantly at $\lambda=$1 $\mu$m compared to that at $\lambda=$0.6 $\mu$m. For the same reason, the total disk albedo at 0.6 $\mu$m is only slightly higher than that at 1 $\mu$m for the cloud-free case which is also suggested by Figure~\ref{fig:reflected}.

The effect of clouds is twofold. For a very low value of surface albedo, as in the case of the gaseous planets, the presence of clouds increases the depolarization of radiations due to multiple scattering as the single-scattering albedo of the atmosphere increases \citep{chakrabarty2021generic}. This causes the total disk polarization to drop with the presence of clouds while the disk albedo rises (see Figures 15-17 of \cite{chakrabarty2021generic}). On the other hand, for a rocky planet with relatively high surface albedo, we find that stream-ii dominates over stream-i at the TOA in the absence of any cloud particle which causes a low value of disk polarization. However, the presence of a cloud layer tends to strengthen stream-i by reflecting back more of the downward radiations to the upward direction and weaken stream-ii by reflecting them back in the downward direction. As a result, the presence of clouds increases the overall disk-integrated degree of polarization and also increases the albedo of the disk. Thus, polarization serves as an indicator of the presence of clouds and helps us understand the thickness and the properties of the cloud layers when combined with the scalar spectrum of the planet.

\section{Conclusions}\label{sec:conc}

This paper focuses on the various existing techniques that can be used synergically to characterize Earth-like exoplanets. We have demonstrated how the inclusion of diffuse radiation due to scattering can improve the model of transmission spectra over the traditional approach of invoking the Beer-Bouguer-lambert's law. The difference is significant with respect to the molecular absorption features that can serve as  {biosignatures}. We have also presented the reflection spectra including non-zero globally averaged surface albedo and these spectra also carry information about the  {biosignatures} and the volatiles. However, obtaining the transmission or reflection spectra from such small-sized planets with thin atmospheres is extremely challenging at present but will be possible in the era of the upcoming big-budget missions like HabEx, LUVOIR, TMT, ELT,  etc. Our models will play a significant role in the habitability study of the Earth-like planets using transmission, reflection spectra and phase dependent linear polarization.

In this paper we demonstrate that atmospheric cloud can significantly affect both the transmission and reflection spectra. 
The use of polarimetry can allow us to study the properties of the clouds in great detail and reduce the overshadowing effects of clouds. The coronagraphic instruments of those upcoming missions in the  {upcoming decades} will be able to directly image the Earth-like planets in the habitable zones around their host stars. Leveraging the polarimetric instruments in conjunction with these coronagraphic instruments, we will be able to conduct a phase curve study of such planets. Our vector phase curve models show the contrast required to resolve these planets from their host stars and also predict the maximum observable reflected flux and degree of polarization.

Evidently, the surface albedo and the clouds significantly dictate the nature of the phase dependent light curves. Our approximate globally averaged Lambertian representation of the surface albedo has allowed us to simplify the calculations to some extent and develop an understanding of the effect of surface albedo on the reflection spectra and phase dependent light curves. However, in our upcoming work, we will consider individual surface components and their wavelength-dependent reflection matrices to calculate the spectra and the light curves more accurately.

Finally, our models should be useful in designing the instruments onboard the upcoming missions, selecting the science targets, as well as extracting the planetary properties from the spectra and phase curves, once obtained.

\vspace{5mm}

\bibliography{main}{}

\begin{thebibliography}{}
\expandafter\ifx\csname natexlab\endcsname\relax\def\natexlab#1{#1}\fi
\providecommand{\url}[1]{\href{#1}{#1}}
\providecommand{\dodoi}[1]{doi:~\href{http://doi.org/#1}{\nolinkurl{#1}}}
\providecommand{\doeprint}[1]{\href{http://ascl.net/#1}{\nolinkurl{http://ascl.net/#1}}}
\providecommand{\doarXiv}[1]{\href{https://arxiv.org/abs/#1}{\nolinkurl{https://arxiv.org/abs/#1}}}

\bibitem[{Batalha {et~al.}(2020)Batalha, Freedman, Lupu, \&
  Marley}]{natasha_batalha_2020_3759675}
Batalha, N., Freedman, R., Lupu, R., \& Marley, M. 2020, Resampled Opacity
  Database for PICASO v2, 1.0,  Zenodo, \dodoi{10.5281/zenodo.3759675}

\bibitem[{Batalha {et~al.}(2019)Batalha, Marley, Lewis, \&
  Fortney}]{batalha2019exoplanet}
Batalha, N.~E., Marley, M.~S., Lewis, N.~K., \& Fortney, J.~J. 2019, The
  Astrophysical Journal, 878, 70

\bibitem[{Berdyugina {et~al.}(2018)Berdyugina, Kuhn, Langlois, Moretto,
  Krissansen-Totton, Catling, Grenfell, Santl-Temkiv, Finster, Tarter,
  {et~al.}}]{berdyugina2018exo}
Berdyugina, S.~V., Kuhn, J.~R., Langlois, M., {et~al.} 2018, in Ground-based
  and Airborne Telescopes VII, Vol. 10700, SPIE, 1453--1466

\bibitem[{Chakrabarty \& Sengupta(2020)}]{chakrabarty2020effects}
Chakrabarty, A., \& Sengupta, S. 2020, The Astrophysical Journal, 898, 89

\bibitem[{Chakrabarty \& Sengupta(2021)}]{chakrabarty2021generic}
---. 2021, The Astrophysical Journal, 917, 83

\bibitem[{Claudi \& Alei(2019)}]{claudi2019biosignatures}
Claudi, R., \& Alei, E. 2019, Galaxies, 7, 82

\bibitem[{Coffeen \& Gehrels(1969)}]{coffeen1969wavelength}
Coffeen, D., \& Gehrels, T. 1969, The Astronomical Journal, 74, 433

\bibitem[{Coffeen(1969)}]{coffeen1969}
Coffeen, D.~L. 1969, The Astronomical Journal, 74, 446

\bibitem[{Covone {et~al.}(2021)Covone, Ienco, Cacciapuoti, \&
  Inno}]{covone2021efficiency}
Covone, G., Ienco, R.~M., Cacciapuoti, L., \& Inno, L. 2021, Monthly Notices of
  the Royal Astronomical Society, 505, 3329

\bibitem[{Ehrenreich {et~al.}(2006)Ehrenreich, Tinetti, Des~Etangs,
  Vidal-Madjar, \& Selsis}]{ehrenreich2006transmission}
Ehrenreich, D., Tinetti, G., Des~Etangs, A.~L., Vidal-Madjar, A., \& Selsis, F.
  2006, Astronomy and Astrophysics, 448, 379

\bibitem[{Fauchez {et~al.}(2017)Fauchez, Rossi, \& Stam}]{fauchez2017o2}
Fauchez, T., Rossi, L., \& Stam, D.~M. 2017, The Astrophysical Journal, 842, 41

\bibitem[{Fujii {et~al.}(2018)Fujii, Angerhausen, Deitrick, Domagal-Goldman,
  Grenfell, Hori, Kane, Pall{\'e}, Rauer, Siegler,
  {et~al.}}]{fujii2018exoplanet}
Fujii, Y., Angerhausen, D., Deitrick, R., {et~al.} 2018, Astrobiology, 18, 739

\bibitem[{Gialluca {et~al.}(2021)Gialluca, Robinson, Rugheimer, \&
  Wunderlich}]{gialluca2021characterizing}
Gialluca, M.~T., Robinson, T.~D., Rugheimer, S., \& Wunderlich, F. 2021,
  Publications of the Astronomical Society of the Pacific, 133, 054401

\bibitem[{Gordon {et~al.}(2022)Gordon, Karalidi, Bott, Mulder, Miles-P{\'a}ez,
  \& Stam}]{gordon2022polarized}
Gordon, K.~E., Karalidi, T., Bott, K.~M., {et~al.} 2022, Bulletin of the
  American Astronomical Society, 54, 102

\bibitem[{Grenfell {et~al.}(2014)Grenfell, Gebauer, Paris, Godolt, \&
  Rauer}]{grenfell2014sensitivity}
Grenfell, J.~L., Gebauer, S., Paris, P.~v., Godolt, M., \& Rauer, H. 2014,
  Planetary and Space Science, 98, 66

\bibitem[{Groot {et~al.}(2020)Groot, Rossi, Trees, Cheung, \&
  Stam}]{groot2020colors}
Groot, A., Rossi, L., Trees, V., Cheung, J., \& Stam, D. 2020, Astronomy \&
  Astrophysics, 640, A121

\bibitem[{Hall \& Riley(1974)}]{hall1974photometric}
Hall, J., \& Riley, L. 1974, Icarus, 23, 144

\bibitem[{Huang(1959)}]{huang1959occurrence}
Huang, S.-S. 1959, American Scientist, 47, 397

\bibitem[{Huang(1960)}]{huang1960life}
---. 1960, Scientific American, 202, 55

\bibitem[{Joos \& Schmid(2007)}]{joos2007limb}
Joos, F., \& Schmid, H. 2007, Astronomy \& Astrophysics, 463, 1201

\bibitem[{Kaltenegger \& Traub(2009)}]{kaltenegger2009transits}
Kaltenegger, L., \& Traub, W.~A. 2009, The Astrophysical Journal, 698, 519

\bibitem[{Kaltenegger {et~al.}(2007)Kaltenegger, Traub, \&
  Jucks}]{kaltenegger2007spectral}
Kaltenegger, L., Traub, W.~A., \& Jucks, K.~W. 2007, The Astrophysical Journal,
  658, 598

\bibitem[{Kane {et~al.}(2016)Kane, Hill, Kasting, Kopparapu, Quintana, Barclay,
  Batalha, Borucki, Ciardi, Haghighipour, {et~al.}}]{kane2016catalog}
Kane, S.~R., Hill, M.~L., Kasting, J.~F., {et~al.} 2016, The Astrophysical
  Journal, 830, 1

\bibitem[{Karalidi \& Stam(2012)}]{karalidi2012modeled}
Karalidi, T., \& Stam, D. 2012, Astronomy \& Astrophysics, 546, A56

\bibitem[{Karalidi {et~al.}(2011)Karalidi, Stam, \&
  Hovenier}]{karalidi2011flux}
Karalidi, T., Stam, D., \& Hovenier, J. 2011, Astronomy \& Astrophysics, 530,
  A69

\bibitem[{Karalidi {et~al.}(2012{\natexlab{a}})Karalidi, Stam, \&
  Hovenier}]{karalidi2012looking}
---. 2012{\natexlab{a}}, Astronomy \& Astrophysics, 548, A90

\bibitem[{Karalidi {et~al.}(2012{\natexlab{b}})Karalidi, Stam, Snik, Bagnulo,
  Sparks, \& Keller}]{karalidi2012observing}
Karalidi, T., Stam, D., Snik, F., {et~al.} 2012{\natexlab{b}}, Planetary and
  Space Science, 74, 202

\bibitem[{Kasting {et~al.}(1993)Kasting, Whitmire, \&
  Reynolds}]{kasting1993habitable}
Kasting, J.~F., Whitmire, D.~P., \& Reynolds, R.~T. 1993, ICARUS, 101, 108

\bibitem[{Kawashima \& Rugheimer(2019)}]{kawashima2019theoretical}
Kawashima, Y., \& Rugheimer, S. 2019, The Astronomical Journal, 157, 213

\bibitem[{Kedziora-Chudczer \& Bailey(2010)}]{kedziora2010using}
Kedziora-Chudczer, L., \& Bailey, J. 2010, in Pathways Towards Habitable
  Planets, Vol. 430, 469

\bibitem[{Kempton {et~al.}(2017)Kempton, Lupu, Owusu-Asare, Slough, \&
  Cale}]{kempton2017exo}
Kempton, E. M.-R., Lupu, R., Owusu-Asare, A., Slough, P., \& Cale, B. 2017,
  Publications of the Astronomical Society of the Pacific, 129, 044402

\bibitem[{Kitzmann {et~al.}(2010{\natexlab{a}})Kitzmann, Patzer, von Paris,
  Godolt, Stracke, Gebauer, Grenfell, \& Rauer}]{kitzmann2010clouds}
Kitzmann, D., Patzer, A., von Paris, P., {et~al.} 2010{\natexlab{a}}, Astronomy
  and Astrophysics, 511, A66

\bibitem[{Kitzmann {et~al.}(2010{\natexlab{b}})Kitzmann, Vasquez, Patzer,
  Schreier, Rauer, \& Trautmann}]{kitzmann2010influence}
Kitzmann, D., Vasquez, M., Patzer, A., {et~al.} 2010{\natexlab{b}}, in European
  Planetary Science Congress 2010, 725

\bibitem[{Klind{\v{z}}i{\'c} {et~al.}(2021)Klind{\v{z}}i{\'c}, Stam, Snik,
  Keller, Hoeijmakers, van Dam, Willebrands, Karalidi, Pallichadath, van Dijk,
  {et~al.}}]{klindvzic2021loupe}
Klind{\v{z}}i{\'c}, D., Stam, D.~M., Snik, F., {et~al.} 2021, Philosophical
  Transactions of the Royal Society A, 379, 20190577

\bibitem[{Kopparapu {et~al.}(2013)Kopparapu, Ramirez, Kasting, Eymet, Robinson,
  Mahadevan, Terrien, Domagal-Goldman, Meadows, \&
  Deshpande}]{kopparapu2013habitable}
Kopparapu, R.~K., Ramirez, R., Kasting, J.~F., {et~al.} 2013, The Astrophysical
  Journal, 765, 131

\bibitem[{Lin {et~al.}(2021)Lin, MacDonald, Kaltenegger, \&
  Wilson}]{lin2021differentiating}
Lin, Z., MacDonald, R.~J., Kaltenegger, L., \& Wilson, D.~J. 2021, Monthly
  Notices of the Royal Astronomical Society, 505, 3562

\bibitem[{Madden \& Kaltenegger(2020)}]{madden2020high}
Madden, J., \& Kaltenegger, L. 2020, The Astrophysical Journal Letters, 898,
  L42

\bibitem[{Mallama(2009)}]{mallama2009characterization}
Mallama, A. 2009, Icarus, 204, 11

\bibitem[{Marley \& Sengupta(2011)}]{marley2011probing}
Marley, M.~S., \& Sengupta, S. 2011, Monthly Notices of the Royal Astronomical
  Society, 417, 2874

\bibitem[{Michael F.~Sterzik \& Manev(2020)}]{michael2020}
Michael F.~Sterzik, Stefano~Bagnulo, C.~E., \& Manev, M. 2020, Astronomy and
  Astrophysics, \dodoi{https://doi.org/10.1051/0004-6361/202038270}

\bibitem[{Michalsky \& Stokes(1977)}]{michalsky1977whole}
Michalsky, J., \& Stokes, R. 1977, The Astrophysical Journal, 213, L135

\bibitem[{Molina {et~al.}(2017)Molina, Stam, \& Rossi}]{berzosa2017traces}
Molina, J., Stam, D., \& Rossi, L. 2017, in European Planetary Science Congress
  2017

\bibitem[{Molina {et~al.}(2018)Molina, Rossi, \& Stam}]{molina2018traces}
Molina, J.~B., Rossi, L., \& Stam, D.~M. 2018, Astronomy \& Astrophysics, 618,
  A162

\bibitem[{Mu{\~n}oz(2018)}]{munoz2018mapping}
Mu{\~n}oz, A.~G. 2018, The Astrophysical Journal, 854, 108

\bibitem[{Owen(1980)}]{owen1980search}
Owen, T. 1980, in Strategies for the Search for Life in the Universe
  (Springer), 177--185

\bibitem[{Pall{\'e} {et~al.}(2010)Pall{\'e}, Mu{\~n}oz, Osorio,
  Monta{\~n}{\'e}s-Rodr{\'\i}guez, Barrena, \&
  Mart{\'\i}n}]{palle2010observations}
Pall{\'e}, E., Mu{\~n}oz, A.~G., Osorio, M. R.~Z., {et~al.} 2010, Proc Int
  Astron Union, 6, 385

\bibitem[{Pall{\'e} {et~al.}(2009)Pall{\'e}, Osorio, Barrena,
  Monta{\~n}{\'e}s-Rodr{\'\i}guez, \& Mart{\'\i}n}]{palle2009earth}
Pall{\'e}, E., Osorio, M. R.~Z., Barrena, R., Monta{\~n}{\'e}s-Rodr{\'\i}guez,
  P., \& Mart{\'\i}n, E.~L. 2009, Nature, 459, 814

\bibitem[{Patty {et~al.}(2021)Patty, K{\"u}hn, Lambrev, Spadaccia, Hoeijmakers,
  Keller, Mulder, Pallichadath, Poch, Snik, {et~al.}}]{patty2021biosignatures}
Patty, C.~L., K{\"u}hn, J.~G., Lambrev, P.~H., {et~al.} 2021, Astronomy \&
  Astrophysics, 651, A68

\bibitem[{Peraiah \& Grant(1973)}]{peraiah1973numerical}
Peraiah, A., \& Grant, I. 1973, J. Int. Math. Appl, 12, 75

\bibitem[{Rice(2014)}]{rice2014detection}
Rice, K. 2014, Challenges, 5, 296

\bibitem[{Rossi {et~al.}(2018)Rossi, Berzosa-Molina, \&
  Stam}]{rossi2018pymiedap}
Rossi, L., Berzosa-Molina, J., \& Stam, D.~M. 2018, Astronomy \& Astrophysics,
  616, A147

\bibitem[{Rossi {et~al.}(2017)Rossi, Stam, \& Turbet}]{rossi2017retrieving}
Rossi, L., Stam, D., \& Turbet, M. 2017, in European Planetary Science Congress
  2017

\bibitem[{Rossi \& Stam(2018)}]{rossi2018circular}
Rossi, L., \& Stam, D.~M. 2018, Astronomy \& Astrophysics, 616, A117

\bibitem[{Rugheimer \& Kaltenegger(2018)}]{rugheimer2018spectra}
Rugheimer, S., \& Kaltenegger, L. 2018, The Astrophysical Journal, 854, 19

\bibitem[{Rugheimer {et~al.}(2013)Rugheimer, Kaltenegger, Zsom, Segura, \&
  Sasselov}]{rugheimer2013spectral}
Rugheimer, S., Kaltenegger, L., Zsom, A., Segura, A., \& Sasselov, D. 2013,
  Astrobiology, 13, 251

\bibitem[{Sagan {et~al.}(1993)Sagan, Thompson, Carlson, Gurnett, \&
  Hord}]{sagan1993search}
Sagan, C., Thompson, W.~R., Carlson, R., Gurnett, D., \& Hord, C. 1993, Nature,
  365, 715

\bibitem[{Scharf(2009)}]{scharf2009extrasolar}
Scharf, C. 2009, Extrasolar planets and astrobiology (Sausalito, Calif:
  University Science Books)

\bibitem[{Seager(2008)}]{seager2008exoplanet}
Seager, S. 2008, Space Science Reviews, 135, 345

\bibitem[{Seager(2010)}]{seager2010exoplanet}
---. 2010, Exoplanet atmospheres : physical processes (Princeton, N.J:
  Princeton University Press)

\bibitem[{Segura {et~al.}(2005)Segura, Kasting, Meadows, Cohen, Scalo, Crisp,
  Butler, \& Tinetti}]{segura2005biosignatures}
Segura, A., Kasting, J.~F., Meadows, V., {et~al.} 2005, Astrobiology, 5, 706

\bibitem[{Segura {et~al.}(2007)Segura, Meadows, Kasting, Crisp, \&
  Cohen}]{segura2007abiotic}
Segura, A., Meadows, V., Kasting, J., Crisp, D., \& Cohen, M. 2007, Astronomy
  \& Astrophysics, 472, 665

\bibitem[{Selsis(2004)}]{selsis2004atmosphericbiomarkers}
Selsis, F. 2004, Boletin SEA, 12, 27

\bibitem[{Selsis {et~al.}(2002)Selsis, Despois, \&
  Parisot}]{selsis2002signature}
Selsis, F., Despois, D., \& Parisot, J.-P. 2002, Astronomy \& Astrophysics,
  388, 985

\bibitem[{Selsis {et~al.}(2008)Selsis, Kaltenegger, \&
  Paillet}]{selsis2008terrestrial}
Selsis, F., Kaltenegger, L., \& Paillet, J. 2008, Physica Scripta, 2008, 014032

\bibitem[{Sengupta(2008)}]{sengupta2008cloudy}
Sengupta, S. 2008, The Astrophysical Journal, 683, L195

\bibitem[{Sengupta(2016)}]{sengupta2016polarimetric}
---. 2016, The Astronomical Journal, 152, 98

\bibitem[{Sengupta(2018)}]{sengupta2018polarization}
---. 2018, The Astrophysical Journal, 861, 41

\bibitem[{Sengupta {et~al.}(2020)Sengupta, Chakrabarty, \&
  Tinetti}]{sengupta2020optical}
Sengupta, S., Chakrabarty, A., \& Tinetti, G. 2020, The Astrophysical Journal,
  889, 181

\bibitem[{Sengupta \& Krishan(2001)}]{sengupta2001probing}
Sengupta, S., \& Krishan, V. 2001, The Astrophysical Journal, 561, L123

\bibitem[{Sengupta \& Maiti(2006)}]{sengupta2006polarization}
Sengupta, S., \& Maiti, M. 2006, The Astrophysical Journal, 639, 1147

\bibitem[{Sengupta \& Marley(2009)}]{sengupta2009multiple}
Sengupta, S., \& Marley, M.~S. 2009, The Astrophysical Journal, 707, 716

\bibitem[{Sengupta \& Marley(2010)}]{sengupta2010observed}
---. 2010, The Astrophysical Journal Letters, 722, L142

\bibitem[{Sengupta \& Marley(2011)}]{sengupta2011multiple}
---. 2011, Pramana, 77, 157

\bibitem[{Sengupta \& Marley(2016)}]{sengupta2016detecting}
---. 2016, The Astrophysical Journal, 824, 76

\bibitem[{Sing {et~al.}(2016)Sing, Fortney, Nikolov, Wakeford, Kataria, Evans,
  Aigrain, Ballester, Burrows, Deming, {et~al.}}]{sing2016continuum}
Sing, D.~K., Fortney, J.~J., Nikolov, N., {et~al.} 2016, Nature, 529, 59

\bibitem[{Stam(2008)}]{stam2008spectropolarimetric}
Stam, D. 2008, Astronomy \& Astrophysics, 482, 989

\bibitem[{Stam {et~al.}(2006)Stam, De~Rooij, Cornet, \&
  Hovenier}]{stam2006integrating}
Stam, D., De~Rooij, W., Cornet, G., \& Hovenier, J. 2006, Astronomy \&
  Astrophysics, 452, 669

\bibitem[{Stam \& Hovenier(2005)}]{stam2005errors}
Stam, D., \& Hovenier, J. 2005, Astronomy \& Astrophysics, 444, 275

\bibitem[{Stam(2003)}]{stam2003polarizationb}
Stam, D.~M. 2003, in Earths: DARWIN/TPF and the Search for Extrasolar
  Terrestrial Planets, Vol. 539, 615--619

\bibitem[{Stam {et~al.}(2003)Stam, Hovenier, \& Waters}]{stam2003polarization}
Stam, D.~M., Hovenier, J., \& Waters, R. 2003, in Scientific Frontiers in
  Research on Extrasolar Planets, Vol. 294, 535--538

\bibitem[{Sterzik {et~al.}(2019)Sterzik, Bagnulo, Stam, Emde, \&
  Manev}]{sterzik2019spectral}
Sterzik, M.~F., Bagnulo, S., Stam, D.~M., Emde, C., \& Manev, M. 2019,
  Astronomy \& Astrophysics, 622, A41

\bibitem[{Tinetti(2006)}]{tinetti2006characterizing}
Tinetti, G. 2006, Origins of life and evolution of biospheres, 36, 541

\bibitem[{Tinetti {et~al.}(2013)Tinetti, Encrenaz, \& Coustenis}]{Tinetti}
Tinetti, G., Encrenaz, T., \& Coustenis, A. 2013, The Astronomy and
  Astrophysics Review, 21, 63, \dodoi{10.1007/s00159-013-0063-6}

\bibitem[{Torres {et~al.}(2015)Torres, Kipping, Fressin, Caldwell, Twicken,
  Ballard, Batalha, Bryson, Ciardi, Henze, {et~al.}}]{torres2015validation}
Torres, G., Kipping, D.~M., Fressin, F., {et~al.} 2015, The Astrophysical
  Journal, 800, 99

\bibitem[{Wang {et~al.}(2019)Wang, Qu, \& Li}]{wang2019diurnal}
Wang, S., Qu, Z.-Q., \& Li, H. 2019, Research in Astronomy and Astrophysics,
  19, 117

\bibitem[{Wei \& Zhong-quan(2017)}]{wei2017polarimetric}
Wei, S., \& Zhong-quan, Q. 2017, Chinese Astronomy and Astrophysics, 41, 235

\bibitem[{West {et~al.}(1983)West, Sato, Hart, Lane, Hord, Simmons, Esposito,
  Coffeen, \& Pomphrey}]{west1983photometry}
West, R., Sato, M., Hart, H., {et~al.} 1983, Journal of Geophysical Research:
  Space Physics, 88, 8679

\bibitem[{Whitmire {et~al.}(1991)Whitmire, Reynolds, \&
  Kasting}]{whitmire1991habitable}
Whitmire, D., Reynolds, R., \& Kasting, J. 1991, in Bioastronomy (Springer),
  173--178

\bibitem[{Wunderlich {et~al.}(2019)Wunderlich, Godolt, Grenfell, St{\"a}dt,
  Smith, Gebauer, Schreier, Hedelt, \& Rauer}]{wunderlich2019detectability}
Wunderlich, F., Godolt, M., Grenfell, J.~L., {et~al.} 2019, Astronomy and
  Astrophysics, 624, A49

\bibitem[{Wunderlich {et~al.}(2020)Wunderlich, Scheucher, Godolt, Grenfell,
  Schreier, Schneider, Wilson, S{\'a}nchez-L{\'o}pez, L{\'o}pez-Puertas, \&
  Rauer}]{wunderlich2020distinguishing}
Wunderlich, F., Scheucher, M., Godolt, M., {et~al.} 2020, The Astrophysical
  Journal, 901, 126

\bibitem[{Yan {et~al.}(2015)Yan, Fosbury, Petr-Gotzens, Zhao, Wang, Wang, Liu,
  \& Pall{\'e}}]{yan2015high}
Yan, F., Fosbury, R.~A., Petr-Gotzens, M.~G., {et~al.} 2015, Int. J.
  Astrobiology, 14, 255

\bibitem[{Zubko {et~al.}(2008)Zubko, Baba, \& Murakami}]{zubko2008polarization}
Zubko, N., Baba, N., \& Murakami, N. 2008, in Space Telescopes and
  Instrumentation 2008: Optical, Infrared, and Millimeter, Vol. 7010, SPIE,
  503--507

\bibitem[{Zugger {et~al.}(2011)Zugger, Kasting, Williams, Kane, \&
  Philbrick}]{zugger2011searching}
Zugger, M., Kasting, J., Williams, D., Kane, T., \& Philbrick, C. 2011, The
  Astrophysical Journal, 739, 12

\bibitem[{Zugger {et~al.}(2010)Zugger, Kasting, Williams, Kane, \&
  Philbrick}]{zugger2010light}
Zugger, M.~E., Kasting, J.~F., Williams, D.~M., Kane, T.~J., \& Philbrick,
  C.~R. 2010, The Astrophysical Journal, 723, 1168

\end{thebibliography}
\bibliographystyle{aasjournal}

\end{document}